\documentstyle[prl,aps,epsf,multicol]{revtex}
\begin{document}
\draft


\title{Coulomb Charging Effects for Finite Channel Number}
\author{Georg G\"oppert$^{1,2}$, Hermann Grabert$^{1}$, 
and Christian Beck$^{3}$} 
\address{
   $^1$Fakult\"at f\"ur Physik, Albert--Ludwigs--Universit\"at, \\
     Hermann--Herder--Stra{\ss}e~3, D--79104 Freiburg, Germany \\
   $^2$Service de Physique de l'Etat Condens\'e, \\
           CEA-Saclay, 91191 Gif-sur-Yvette, France \\
   $^3$Centrum f\"ur Complexe Systeme und Visualisierung, \\
           Universit\"atsallee 29, D--28359 Bremen}



\maketitle
\widetext

\begin{abstract}
We consider quantum fluctuations of the charge on a small metallic 
grain caused by virtual electron tunneling to a nearby electrode. The 
average electron number and the effective charging energy are
determined by means of
perturbation theory in the tunneling Hamiltonian. In particular we
discuss the dependence of charging effects on the number $N$ of 
tunneling channels. Earlier results for $N \gg 1$ are found to be
approached rather rapidly with increasing $N$.
\end{abstract}


Single electron effects are well studied for metallic junctions in the 
region of weak tunneling \cite{orthodox}. In the case of a large
tunneling conductance $G_T$ comparable or even larger than the conductance
quantum $G_K=e^2/h$, it is necessary to go beyond lowest order
perturbation theory in the tunneling Hamiltonian \cite{Matveev} - 
\nocite{physica, Grabertprb, zaikin, WangSluggon, Rapid, zwerger, koenig} 
\cite{georgprl}. In this case
the theory involves contributions of order $1/N$ where $N$ is the 
number of transport channels. Since the
junction area is typically much larger than $\lambda_F^2$, where
$\lambda_F$ is the Fermi wavelength, the number of tunneling 
channels $N$ is often very large for metallic junctions, typically about
$10^4$. Terms of order $1/N$ are thus dropped in most former
approaches. In contrast
to lithographically fabricated metallic junctions, in break
junctions \cite{break} only a
small number of channels may be available and terms proportional to
$1/N$  cannot
be neglected. 
Also for tunnel barriers \cite{barrier} in a two-dimensional
electron gas there are typically only $2$ spin degenerate 
transport channels contributing to the tunneling
current. Thus, it is interesting to investigate 
how Coulomb blockade effects are modified by
such $1/N$ corrections. Here we focus on systems where the
dimensionless conductance per channel, $g_0=G_T/4 \pi^2 N G_K$,
remains small, 
that is to cases where the channels contributing to 
charge transfer are weakly transmitting. Further we
restrict ourselves to the zero temperature case.
This covers only partly the range of experimental interest but the results 
indicate how relevant $1/N$ corrections are.

\begin{figure}
\begin{center}
\leavevmode
\epsfxsize=0.6 \textwidth
\epsfbox{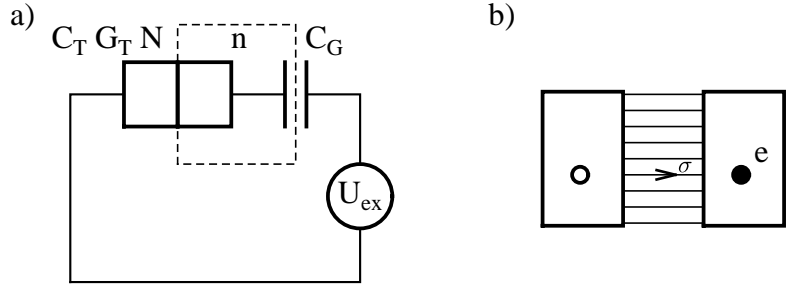}
\end{center}
\caption{a) Circuit diagram of a metallic grain (area within dashed
line) coupled by a tunnel junction to the left electrode and
capacitively to the right electrode. b)
Electron-hole pair excitation in channel $\sigma$ of the tunnel junction.}
\label{fig:fig1}
\end{figure}

Specifically, we consider a small metallic grain in between two
bulk electrodes biased by a voltage source $U_{\rm ex}$. 
The circuit diagram of the system is depicted in
Fig.~\ref{fig:fig1}a). Between the grain and the left bulk electrode
electrons can tunnel while the right electrode
couples purely capacitively. Thus no
dc-current flows through the system. Due to the discreteness of the
tunneling process, the excess charge of the grain can be shifted only
by multiples of the elementary charge $e$, and we introduce an excess
charge number $n$ which characterizes the island charge
$q=-ne$. As far as the tunneling conductance per channel is small
compared to the conductance quantum $G_K$, electron tunneling can be
described in terms
of a tunneling Hamiltonian which will be treated as a perturbation.
The unperturbed Hamiltonian for the Fermi liquids and the Coulomb
energy reads
\begin{equation}
  H_0=E_c(n-n_{\rm ex})^2
  + \sum_{k\sigma} \epsilon_{k\sigma} a^\dagger_{k\sigma} a_{k\sigma}
  + \sum_{q\sigma} \epsilon_{q\sigma} a^\dagger_{q\sigma} a_{q\sigma},
\label{eq:ungesthamilonian}
\end{equation}
where $n_{\rm ex}=C_G U_{\rm ex}/e$ is a dimensionless voltage and 
\begin{equation}
E_c=\frac{e^2}{2(C_T+C_G)}
\end{equation}
is the charging energy needed to transfer one electron to the uncharged
island at $n_{\rm ex}=0$. $\epsilon_{k\sigma}$ and $\epsilon_{q\sigma}$
are one-particle-energies for channel index $\sigma$ and longitudinal
quantum number $k$ and $q$, respectively. The operator
$a^\dagger_{k\sigma}$ ($a_{k\sigma}$) describes creation
(annihilation) of a quasiparticle with quantum
numbers $k\sigma$, and $a^\dagger_{q\sigma}$,
$a_{q\sigma}$ are defined likewise. With these operators the tunneling
Hamiltonian may
be written as
\begin{equation}
  H_T=\sum_{kq\sigma}
    \left( 
      t_{\sigma} a^\dagger_{k\sigma} a_{q\sigma}\Lambda +\mbox{h.c.}
    \right)
\label{eq:tunnelhamiltonian}
\end{equation}
where $t_{\sigma}$ is the tunneling matrix element and $\Lambda$
shifts the island charge number $n$ thereby changing the Coulomb energy. 

\begin{figure}
\begin{center}
\leavevmode
\epsfxsize=0.6 \textwidth
\epsfbox{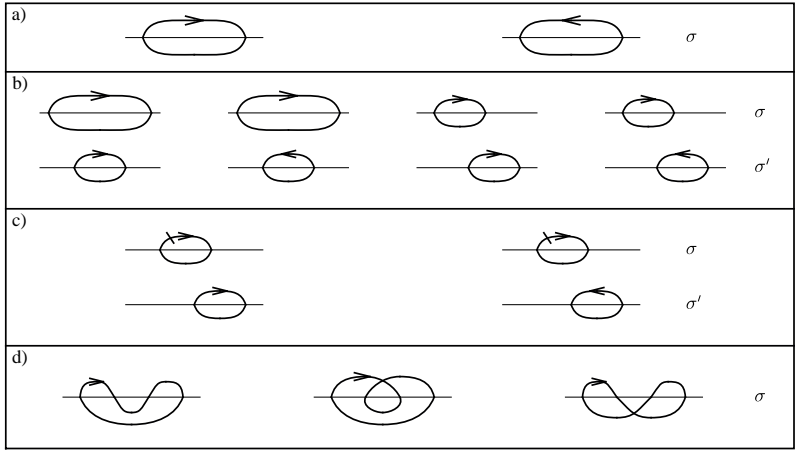}
\end{center}
\caption{Graphical representation of Rayleigh-Schr\"odinger
perturbation theory. a) Graphs of second order. b), c), and d) 
Graphs of fourth order where graphs with inverted arrows are omitted.}
\label{fig:fig2}
\end{figure}

At zero temperature the system is described by the ground state energy
${\cal E}$ and in view of eq.\ $(\ref{eq:ungesthamilonian})$ the
expectation value of the island charge may be written
\begin{equation}
  \langle n \rangle 
   =
  n_{\rm ex}-\frac{1}{2E_c} \frac{\partial {\cal E}}{\partial n_{\rm ex}}.
\end{equation} 
To calculate ${\cal E}$ we make use of Rayleigh-Schr\"odinger
perturbation theory.
The perturbation $(\ref{eq:tunnelhamiltonian})$ contains products of 
one creation and one annihilation operator, thus only even powers in
$t_\sigma$ contribute to ${\cal E}$, and it may be written
\begin{equation}
  {\cal E}=
  {\cal E}^{(0)}+{\cal E}^{(2)}+{\cal E}^{(4)}+{\cal O}(t_\sigma^6).
\label{eq:formalenergie}
\end{equation}
The zeroth order term ${\cal E}^{(0)}$ without tunneling is given by the
minimum of the electrostatic energy and reads $E_c(n_0-n_{\rm ex})^2$
where $n_0$ is the integer closest to $n_{\rm ex}$. Hence, the averaged
island charge in zeroth order perturbation theory is given by the well
known Coulomb staircase. ${\cal E}$ depends on
$n_{\rm ex}$ only via the electrostatic energy. It
is thus an antisymmetric and quasiperiodic function of $n_{\rm ex}$
which allows us to confine ourselves to $0\leq n_{\rm ex}<{\scriptstyle
\frac{1}{2}}$.
For the second order term we get formally
\begin{equation}
  {\cal E}^{(2)}=\langle 0 | H_T Q H_T | 0 \rangle
\label{eq:zweiteoperator}
\end{equation}
with the auxiliary operator
\begin{equation}
  Q=\frac{1-| 0 \rangle \langle 0 |}{{\cal E}_0-H_0}.
\end{equation}
In the same way the fourth order term reads
\begin{equation}
  {\cal E}^{(4)}
  =
  \langle 0 | H_T Q H_T Q H_T Q H_T | 0 \rangle
  -\langle 0 | H_T Q^2 H_T | 0 \rangle \langle 0 | H_T Q H_T | 0 \rangle
\label{eq:vierteoperator}
\end{equation}
where terms with the energy denominator squared arise from the
normalization of the ground state wave function. Inserting the
tunneling Hamiltonian $(\ref{eq:tunnelhamiltonian})$ into
the second order contribution $(\ref{eq:zweiteoperator})$ 
one gets
\begin{equation}
  {\cal E}^{(2)}
  =
  - \sum_{kq\sigma} t_\sigma^2
  \left[
    \frac{\Theta(-\epsilon_{q\sigma}) \Theta(\epsilon_{k\sigma})}
         {\delta E_1+\epsilon_{k\sigma}-\epsilon_{q\sigma}}+
    \frac{\Theta(\epsilon_{q\sigma}) \Theta(-\epsilon_{k\sigma})}
         {\delta E_{-1}-\epsilon_{k\sigma}+\epsilon_{q\sigma}}
  \right]
\label{eq:zweitesumme}
\end{equation}
with the Coulomb energy differences 
$\delta E_{n}=E_c(n^2-2nn_{\rm ex})$. Both summands correspond to the
virtual creation of an electron-hole pair with electron and hole
sitting on  different electrodes, {\it cf}.\ Fig.\ \ref{fig:fig1}b).
This can be represented in terms of Goldstone graphs depicted in
Fig.\ \ref{fig:fig2}. The contributions of second order correspond to
the diagrams in Fig.\ \ref{fig:fig2}a) where the
upper arc with an arrow to the right (left) electrode 
represents the creation of
an electron-hole pair with the electron sitting on the right (left)
and the hole sitting on the other
electrode of the junction. The lower arc destroys this pair and we may
omit the arrow since the process is uniquely
determined by the upper one. 

If we assume that the one particle energy is
separable into a longitudinal and channel part, the summand in eq.\
$(\ref{eq:zweitesumme})$ is independent of the channel $\sigma$ apart
from the factor $t_\sigma^2$. The sum over the transversal and
spin quantum numbers leads to an overall factor $N$ multiplied by the
average $t^2=\sum_\sigma t_\sigma^2/N$ of the tunneling matrix elements. 
We assume the bandwidth $D$ to be large compared to
$E_c$ and take the limit $D/E_c\rightarrow \infty$ at the end of the 
calculation. To keep the finite bandwidth in intermediate formulas, we
introduce an exponential cutoff and replace the sum over the
longitudinal quantum numbers by 
\begin{equation}
  \sum_k F(\epsilon_{k\sigma}) \longrightarrow 
  \rho \int_{-\infty}^\infty 
  {\rm d}\epsilon e^{-|\epsilon|/D} F(\epsilon)
\end{equation}
where $\rho$ is the density of states at the Fermi level. 
Analogously, we introduce an integral with $\rho'$ and $D$ for the $q$
sum. Formula $(\ref{eq:zweitesumme})$ then takes the form
\begin{equation}
  {\cal E}^{(2)}
  =
  -t^2 \rho \rho' N 
  \int_0^\infty {\rm d}\epsilon \, \epsilon \, e^{-|\epsilon|/D}
  \left(
  \frac{1}{\delta E_1+\epsilon} +
  \frac{1}{\delta E_{-1}+\epsilon} 
  \right).
\label{eq:zweiteintegral}
\end{equation}
We now define the dimensionless tunneling conductance per channel by
$g_0=t^2 \rho \rho'$. Whereas the
integral $(\ref{eq:zweiteintegral})$ is divergent for $D \rightarrow
\infty$, the average
island charge to first order in $g_0$ remains finite leading to
\begin{equation}
  \langle n \rangle
  = 
  g_0 N \ln\left(\frac{1+2n_{\rm ex}}{1-2n_{\rm ex}}\right)
  +{\cal O}(g^2)
\end{equation} 
in accordance with previous results \cite{Matveev,physica}.

\begin{figure}
\begin{center}
\leavevmode
\epsfxsize=0.48 \textwidth
\epsfbox{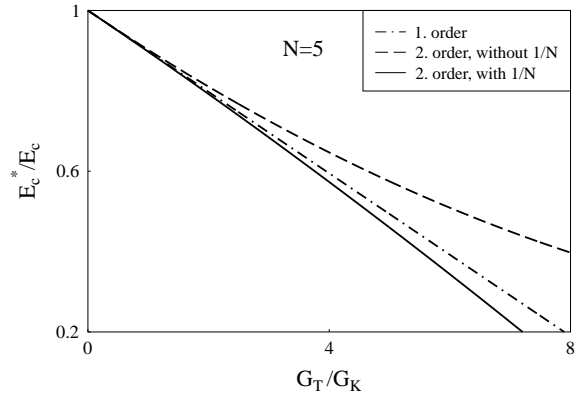}
\end{center}
\caption{Effective charging energy for channel number $N=5$ as a
function of the dimensionless tunneling conductance $G_T/G_K$ in first
and second order perturbation theory with and without $1/N$ corrections.}
\label{fig:fig3}
\end{figure}

Along these lines the fourth order term $(\ref{eq:vierteoperator})$ is
given in terms of double sums over $kq\sigma$. These contributions can
also be represented graphically as twofold virtual electron-hole
pair creation and annihilation. We distinguish three groups of
graphs depicted in Figs.\ \ref{fig:fig2}b), c) and d). The first
set of the graphs represents virtual
electron hole pair creation and annihilation in two distinct channels
$\sigma$ and $\sigma'$ which are only coupled by the energy
denominator. Fig.\ \ref{fig:fig2}c) represents graphs arising from
the normalization of the wave function where the dash signs that the energy
denominator is squared, {\it cf}.\ eq.\ $(\ref{eq:vierteoperator})$. If
there were no Coulomb interaction, i.e. $E_c = 0$, the graphs b) and c)
would cancel in accordance with
the linked cluster theorem for
uncorrelated fermions. Here we have to take
these terms into
account since the Coulomb energy correlates the electrons in the two
electrodes.  The channels
of the graphs in b) and c) are
not restricted and we get by summation over 
the transversal quantum numbers a
factor $N^2$. On the other hand, the graphs in
Fig\ \ref{fig:fig2}d) describe processes within one channel because the
electron created recombines not with the hole created at the same time
but with a different
one which has to have the same channel number. Thus we get by summation over
the channels only a factor $N$. Therefore, we write the average
island charge in the form 
\begin{equation}
 \langle n \rangle
 =
 g_0 N \ln\left(\frac{1+2n_{\rm ex}}{1-2n_{\rm ex}}\right) 
 +(g_0 N)^2 [c(n_{\rm ex})-c(-n_{\rm ex})]
 + g_0^2 N [d(n_{\rm ex})-d(-n_{\rm ex})]
 +{\cal O}(g_0^3).
\label{eq:ergformal}
\end{equation}
Previous theories for $N \gg 1$ introduce the dimensionless
conductance $g=g_0 N$ of the junction and the terms proportional to
$g_0^2 N = g/N$ are dropped.

\begin{figure}
\begin{center}
\leavevmode
\epsfxsize=0.48 \textwidth
\epsfbox{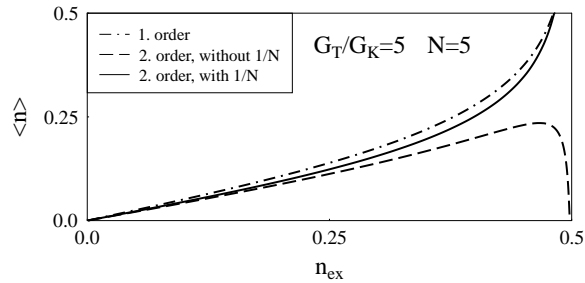}
\end{center}
\caption{The average island charge number for dimensionless tunneling
conductance $G_T/G_K=5$ and channel number $N=5$
as a function of the dimensionless gate voltage $n_{\rm ex}$ .}
\label{fig:fig4}
\end{figure}

The graphs in Fig.\ \ref{fig:fig2} b) and c) do not include 
$1/N$-corrections and they lead to the known second order result
for the island charge \cite{physica}
\begin{eqnarray}
 \,\, c(u) \hspace{-.3cm}
&=&  \hspace{-.3cm}
 -u\left[\frac{4\pi^2}{3}
 +\ln^2\left(\frac{1-2u}{1+2u}\right)\right]    
 - \frac{16(1+2u-2u^2)}
 {(3-2u)(1+2u)} 
 \ln(1-2u)  
                     \nonumber  \\
&&  \hspace{-.3cm}
 - 2(1-u)\left\{ 
   \ln^2 \left[\frac{1-2u}{4(1-u)}\right]
   + 2 \mbox{Li}_2 \left[\frac{3-2u}{4(1-u)}\right] 
     -\frac{8(1-u)}{(1-2u)(3-2u)}
   \ln[4(1-u)] 
  \right\}.     \quad
\label{eq:zweiteN2}
\end{eqnarray}
The graphs in Fig.\ \ref{fig:fig2} d) can also be integrated out
analytically leading to a new contribution of order $1/N$. We find

\begin{eqnarray}
 d(u)
&=&
 \frac{8}{3}\ln^3 \left( \frac{4-4u}{1-2u} \right)
 -\frac{26}{3}\ln^3(1-2u)+15 \ln^3(3-2u)
                     \nonumber  \\
&& 
 +[15\ln(2)+16\ln(4-4u)] \ln^2 \left( \frac{1-2u}{3-2u} \right)
 +[2\ln(1+2u)-7\ln(3-2u)]\ln^2(1-2u)
                     \nonumber  \\
&& 
 +\left[
    10\ln^2 \left(\frac{4-4u}{1-2u} \right)
    -\frac{10 \pi^2}{3}
    +38 \ln^2(3-2u)
  \right]
  \ln \left( \frac{1-2u}{3-2u} \right)
  +\frac{4}{3}\pi^2\ln\left(\frac{3-2u}{4-4u}\right)
                     \nonumber  \\
&& 
 +4\ln\left(\frac{1-2u}{3-2u}\right)
 \left\{
  3\,{\rm Li}_2\left(\frac{1-2u}{3-2u}\right) +
  3\,{\rm Li}_2\left(\frac{3-2u}{4-4u}\right) +  
  2\,{\rm Li}_2\left[\frac{8(1-u)}{(3-2u)^2}\right]
 \right\}
                     \nonumber  \\
&& 
 +6\,{\rm Li}_3\left(\frac{1-2u}{3-2u}\right) 
 -8\,{\rm Li}_3\left(\frac{3-2u}{4-4u}\right) 
 -8\,{\rm Li}_3\left(\frac{1-2u}{4-4u}\right)                   
\label{eq:zweiteN1}
\end{eqnarray}
where ${\rm Li}_2(z)$ is the dilogarithm and ${\rm Li}_3(z)$ the
trilogarithm function \cite{Lewin}. Within second order perturbation theory
the result (\ref{eq:ergformal}) is valid for arbitrary channel numbers 
including $N=1$.

We now compare our result to earlier findings. One of the mostly frequently
discussed quantities is the effective charging energy
\cite{WangSluggon}
$E_c^*$ characterizing the effective strength of the Coulomb blockade
effect.  This quantity is defined by 
\begin{equation}
  E_c^*/E_c=
   \left.
    \frac{1}{2} \frac{\partial^2 {\cal E}}{\partial n_{\rm ex}^2}
   \right|_{n_{\rm ex}=0}.
\end{equation}
For small tunneling conductance, $g_0 \rightarrow 0$, the effective
charging energy approaches the bare charging energy $ E_c$ whereas for
strong electron tunneling $E_c^*$ vanishes. Our analytic
expression leads to
\begin{equation}
  E_c^*/E_c=1 - 4g + 5.066 \ldots g^2 - 7.167 \ldots g^2/N
\end{equation}
where the constants are analytically known but too lengthy to present
here. In Fig.\ \ref{fig:fig3} we show the normalized effective
charging energy in first and second order perturbation theory
without $1/N$ corrections compared to the complete second order
result as a function of the dimensionless conductance $G_T/G_K$. We
see that the $1/N$ corrections become more significant for larger
tunneling conductance.

\begin{figure}
\begin{center}
\leavevmode
\epsfxsize=0.4 \textwidth
\epsfbox{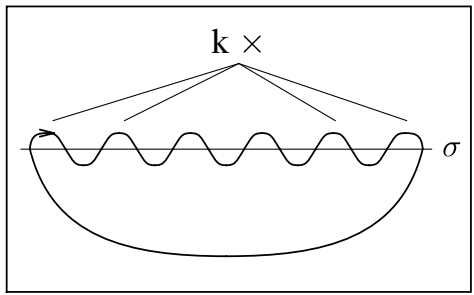}
\end{center}
\caption{Goldstone graph of order $g_0^{k}$ which gives the
leading asymptotic contribution for $n_{\rm ex} \rightarrow 
{\scriptstyle \frac{1}{2}}$.}
\label{fig:fig5}
\end{figure}

In Fig.\ \ref{fig:fig4} the
average island charge $\langle n \rangle$ in first and
second order in $g_0$ with and without the $1/N$ corrections is
depicted for the case  $N=5$ and $G_T/G_K=5$. We see that the
$1/N$ corrections become significant especially for larger external
voltages. 
From a comparison with Monte Carlo data \cite{georgprl} one estimates 
that results of second order perturbation theory are reliable for 
$n_{\rm ex}$ up to $0.3$. In the vicinity of the step, {\it i.e.}, for 
$n_{\rm ex} \rightarrow {\scriptstyle \frac{1}{2}}$, finite order 
perturbation theory diverges and one has to sum diagrams 
of all order \cite{Matveev}. From our {\it eq}.\ $(\ref{eq:zweiteN1})$ 
one sees that 
the $1/N$-corrections become relevant near 
$n_{\rm ex} = {\scriptstyle \frac{1}{2}}$ even for large $N$ since
the qualitative behavior
is changed. For $n_{\rm ex} \rightarrow {\scriptstyle \frac{1}{2}}$,
the leading $N$ terms of second order show the logarithmic
divergence \cite{physica} 
$2(g_0 N)^2\ln^2\delta$ where 
$\delta = {\scriptstyle \frac{1}{2}}-n_{\rm ex}$, whereas the $1/N$
corrections lead to 
$\scriptsize -\frac{4}{3}g_0^2N \ln^3 \delta$ which
eventually dominates the asymptotic
behavior. For the special case $N=1$, this divergence is
in accordance with earlier
findings by Matveev \cite{Matveev}. In general,
for the finite $N$ corrections of order $g_0^k$ we get the leading
asymptotic behavior 
\begin{equation}
 \langle n \rangle
 \sim
 g_0^kN \ln^{2k-1}\delta + {\cal O}(g^{k+1})
\end{equation}
arising from the diagram depicted in Fig.\ \ref{fig:fig5}. These terms
dominate the asymptotics for $n_{\rm ex} \rightarrow 
{\scriptstyle \frac{1}{2}}$, even for large $N$.

In summary, we have studied the influence of finite channel numbers on
the Coulomb staircase. We found that for
small voltages, $n_{\rm ex} \approx  0$, the
$1/N$ corrections are significant up to $N \approx 6$ and become
increasingly important as $n_{\rm ex} = 
{\scriptstyle \frac{1}{2}}$ is approached. While near the step finite
order perturbative results as derived here are not sufficient, the
expression obtained for the effective charging energy is valid within an
experimentally relevant range of parameters.

This work was supported by the Deutsche
Forschungsgemeinschaft (Bonn) and the Deutscher Akademischer Austauschdienst
(DAAD).

\end{document}